\documentclass{elsart}
\usepackage{amsfonts}
\usepackage{amsmath}
\usepackage{graphicx}

\input{tcilatex}

\begin{document}

\begin{frontmatter}%

\title{A Consistent Picture of the Charge Distribution in Reduced Ceria Phases}%
\date{13 November 2009}
\author{E. Shoko, M. F. Smith, Ross H. McKenzie}%

\collab{}%

\address{The University of Queensland, Department of Physics, Brisbane, QLD 4072, Australia}%

\begin{abstract}
We consider the implications of the bond valence model (BVM) description of charge distribution in reduced ceria phases (CeO$_{2-x}$)\cite{Shoko2009,Shoko2009b} to the models used to describe electronic and ionic conductivity in these phases. We conclude that the BVM is consistent with both the small polaron model (SPM) and the atomistic models which describe the electronic and ionic conductivities respectively. For intermediate phases, i.e., $x \sim 0.3$, we suggest the possibility of low temperature metallic conductivity. This has not yet been experimentally observed. We contrast the BVM results and the conventional description of charge distribution in reduced ceria phases.
\end{abstract}%


\end{frontmatter}%

\section*{Introduction}

\label{Sec I} \thispagestyle{plain}Ceria is a technologically important
material with applications in high temperature electrochemical devices \cite%
{Singhal2003,Inaba1996,Steele2000,Kharton2001,Zhang2006}, catalysis \cite%
{Trovarelli2002,Blank2007,Chen2007,Mo2005} and oxygen gas sensors \cite%
{Fray2001,Jasinski2003}. It is the capacity of ceria to rapidly take up and
release oxygen that makes it valuable for these applications.
The process can be described by a reversible chemical reaction:
\begin{equation}
\text{CeO}_{2}\rightleftharpoons \text{CeO}_{2-x}+\frac{x}{2}\text{O}_{2}%
\text{ (g), \ \ \ \ }0\leq x\leq 0.5  \label{reduce}
\end{equation}%
The oxygen exchange controls both the electronic and ionic conductivities in reduced phases, CeO$_{2-x}$,
and both channels of charge transport are important in typical applications. This is illustrated by the example
of a solid-oxide fuel cell where ceria forms the anode. The oxidation of the fuel takes place on the anode and involves the
abstraction of lattice O from the ceria. When an O atom is removed, a
vacancy site results and two extra electrons are left in the crystal lattice
of the anode. For the fuel cell to operate, two processes must ensue. First, the electrons move to the cathode through an external
circuit, which relies on the electronic conductivity of the ceria.
Second, oxide ions migrate from the bulk to the surface of the anode where
the chemical reaction occurs, which requires a sufficiently large ionic conductivity.

A description of the charge distribution in stable structures
of CeO$_{2-x}$ may elucidate the microscopic
mechanism of both the electronic and ionic conductivities. For a long time, it was widely held that when an O vacancy forms
in bulk ceria, the associated charge localizes on two of the nearest neighbour Ce sites
\cite{Martin1974,Skorodumova2002,Fabris2005,Silva2007a,Andersson2007}. We
will call this the standard picture. Recent results from atomistic
simulations challenge this view, indicating that dopant ions prefer to occupy the second or third coordination shell of
the O vacancy \cite%
{Castleton2007,ganduglia-pirovano:026101,li:193401,burow:174710}. These reports are
consistent with our own results from the BVM for slightly reduced ceria \cite%
{Shoko2009,Shoko2009b}, which also contradict the standard picture.

\begin{figure}
[ptb]
\begin{center}
\includegraphics[width=0.9\textwidth]{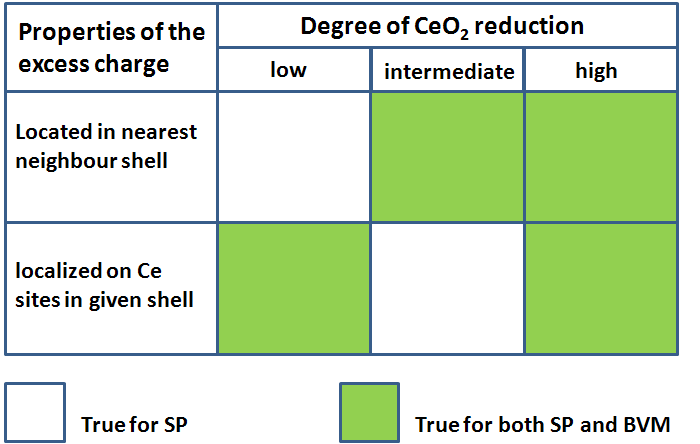}%
\caption{A comparison between the
standard picture and the bond valence model in describing the charge
distribution in reduced ceria phases, CeO$_{2-x}$. In this schematic, we
define `low' as O vacancy concentrations up to just over $x=0.2$,
intermediate for $x\sim 0.3$ and high for $x\sim 0.5$. We compare the two
descriptions with respect to where the excess charge is located and whether
that charge is localized or delocalized. The main difference is that in the
standard picture, the excess charge is localized in the first coordination
shell for all compositions, CeO$_{2-x}$. This picture agrees with the BVM analysis
only in the high composition range. In the low composition range, the excess
charge localizes in the second coordination shell. In the intermediate
region, the BVM analysis implies a delocalized excess charge located in the first
coordination shell.}%
\label{spandbvm}%
\end{center}
\end{figure}

In Fig. \ref{spandbvm}, we summarize the results of the standard picture and the
BVM. %
In the
standard picture, the two extra electrons \emph{localize} in the \emph{%
nearest} neighbour shell for \emph{all} compositions of reduced ceria. The
results of the BVM are different for different composition ranges of the
reduced ceria phases (i.e. for different ranges of $x$ in CeO$_{2-x}$). We roughly define `low',
`intermediate' and `high' $O$-vacancy concentration ranges according to $x\leq 0.2$,
$0.2\leq x\leq 0.3$ and $x\sim 0.5$, respectively. In the low
range, the BVM predicts that the two extra electrons \emph{localize} in the
\emph{next nearest} neighbour shell; in the intermediate region, they \emph{%
delocalize }in the \emph{nearest} neighbour shell while in the high
region, the result is the same as that for the standard picture.

Except for fully reduced ceria, the standard picture and the BVM give qualitatively different
descriptions of charge distribution in CeO$_{2-x}$ phases. We aim to determine whether the
difference in charge distributions results in a corresponding difference in the transport properties of these
materials.  We thus study the electronic and ionic conductivities of the material using appropriate models of charge transport, and ask how the charge distribution will affect model predictions.

In Section \ref{Sec Theory} we provide theoretical background on electronic and ionic conductivity in doped ceria.  We apply these results to study experimental data on the electronic conductivity in the low and intermediate reduction regimes in Sections \ref{Sec II} and \ref{Sec III}, respectively.  The ionic conductivity is examined in Section \ref{Sec IV}.  In Section \ref%
{Sec VI} we conclude.

\section{Theory}
\label{Sec Theory}
Towards understanding general properties of electronic conductivity in reduced ceria we start with a review of the electronic band structures of CeO$_{2}$ and Ce$_{2}$O$_{3}$, which correspond to extremal oxidation states of the Ce oxides.  We then go on to describe the phenomenology of incoherent electron mobility in CeO$_{2-x}$ phases and the relationship between carrier and electronic structure.  It turns out that the ionic mobility, described as a process of O vacancy migration, has similar properties to incoherent electron mobility.  Thus, an important question is how the two types of charge carriers might be distinguished experimentally.
\subsection{Electronic Band Structures of CeO$_{2}$ and Ce$_{2}$O$_{3}$}
\begin{figure}
[ptb]
\begin{center}
\includegraphics[width=0.6\textwidth]{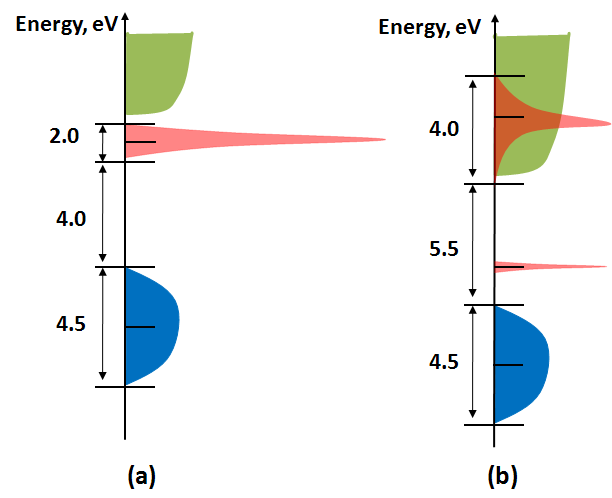}%
\caption{A schematic of the band structures of (a) CeO$_{2}$ and (b) Ce$_{2}$O$_{3}$. Both CeO$_{2}$ and Ce$_{2}$O$_{3}$ are insulators. The band gaps and bandwidths are approximate and estimated from references \cite{Jiang2009,Wuilloud1984,Hay2006,Skorodumova2001,Shoko2009a} for CeO$_{2}$ and references \cite{Nakano1987a,Jiang2009,Shoko2009a} for Ce$_{2}$O$_{3}$. In each case, the bottom band, shaded black (blue in colour) is the valence band which is predominantly O $2p$ in character. The Ce $4f$ states are shown in light grey (red). In the case of CeO$_{2}$, these states are very narrow and centered $\sim 6\unit{\text{eV}}$ above the valence band. For Ce$_{2}$O$_{3}$, the $4f$-manifold splits up into two sectors, a low energy and a high energy sector. The low energy sector is singly occupied and consists of two states per Ce atom while the twelve states in the high energy sector are all unoccupied. In both CeO$_{2}$ and Ce$_{2}$O$_{3}$, the conduction band is shown in grey (green) and are mainly of Ce $5d$ character. When one considers these band structures of CeO$_{2}$ and Ce$_{2}$O$_{3}$, it is not obvious that there could exist an intermediate phase such as Ce$_{7}$O$_{12}$, where low temperature metallic conductivity might occur.}%
\label{bandstructures}%
\end{center}
\end{figure}

CeO$_{2}$ and Ce$_{2}$O$_{3}$ are insulators for entirely different reasons. As can be seen from Fig. \ref{bandstructures}(a), CeO$_{2}$ is a band insulator with a band gap of $\sim 6\unit{eV}$. The valence band is of mainly O $2p$ character while the conduction band consists predominantly of the Ce $5d$ states.  Since there is a negligible probability for electrons to be thermally excited from the valence band to Ce $5d$ states at relevant temperatures, there is no electronic conductivity in stoichiometric CeO$_{2}$. In slightly reduced ceria, electronic conductivity involves the $4f$ states.  The latter exist within a narrow band and thus describe electrons that are approximately localized on individual Ce sites.  Electron hopping between these sites involves a phonon-assisted mechanism.   This process is thermally activated at temperatures below 1000 K or so (a value typical of working device conditions).

In contrast, Ce$_{2}$O$_{3}$ is a Mott insulator \cite{Skorodumova2001,Jiang2009}. The Ce $4f$ states are partially occupied with one electron per site. If the Coulomb repulsion between electrons was taken to be weak, then band theory would predict the system to be a metal. However, the compact $f$ states have a high on-site Coulomb repulsion, $U \sim 6\unit{eV}$, which prohibits the double occupancy of a site. Since electronic conductivity would require that an electron hops onto a Ce site which is already occupied, this material is an insulator with a band gap related to $U$. Fig. \ref{bandstructures} shows that the band gap between the filled valence band and the conduction band is $\sim 5.5\unit{eV}$ and that between the partially occupied Ce $4f$ states and the conduction band is $\sim 3\unit{eV}$. Since $U$ is larger than this energy gap, it is more likely for the electrons to be thermally excited into the conduction band than doubly occupying the local $f$ orbitals. In any case, the number of free carriers is small at relevant temperatures. At low temperatures, the small polaron model (SPM) has been used to describe electronic conductivity in reduced ceria phases \cite{Tuller1977}. We now briefly review the main features of this model.

\subsection{The Phenomenology of Charge Carrier Mobility in CeO$_{2-x}$ Phases}
\label{phenomenon}
The electronic mobility can be equivalently described within two theoretical formalisms: the SPM of Holstein \cite{Holstein1959,Emin1969,Tuller1977} and the Marcus-Hush theory \cite{Marcus1985,Bolton1991}. To model the ionic mobility, we consider O vacancy migration or O self-diffusion in the CeO$_{2-x}$ phases  \cite{Goodenough1984,Morgensen2000,Mehrer2007}.

The fact that electronic mobility in reduced ceria is activated means that the mathematical description of electronic and ionic charge transport is similar (unlike in a metal, where there is no activation barrier to electronic mobility).  We take advantage of this correspondence below, while being careful to point out the significant differences between electronic and ionic transport that remain.

\begin{figure}
[ptb]
\begin{center}
\includegraphics[width=0.99\textwidth]{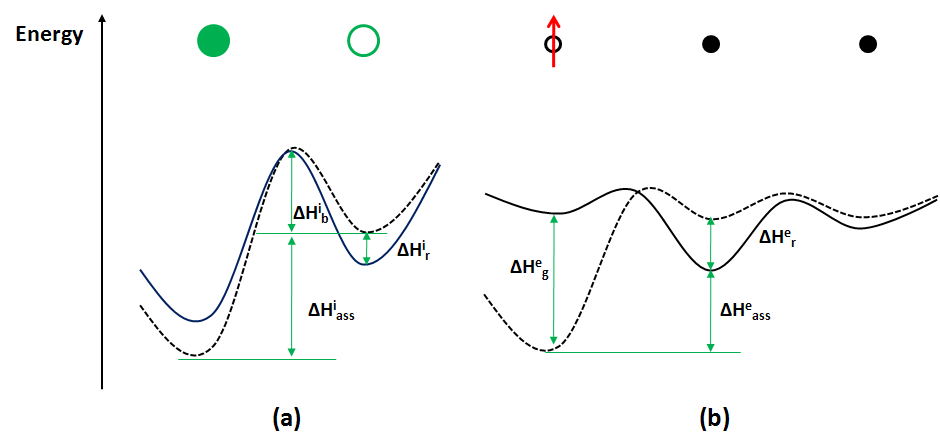}%
\caption{A schematic of the phenomenological description of the mobility of charge carriers in CeO$_{2-x}$ phases. Illustration (a) shows the energies relevant to ionic conductivity. The shaded and empty circles represent an O ion on an O site and an O vacancy respectively. The two sites are both energetically and crystallographically inequivalent. The occupied site is lower in energy by $\Delta H^{i}_{ass}$. As the O ion moves to occupy the vacancy site, it experiences a potential shown by the dotted curve. There is a barrier potential given by $\Delta H^{i}_{b}$ which arises from the saddle point created by the layer of Ce atoms which the O ion has to cross to reach the O vacancy site. The solid curve shows how the potential changes as a result of the lattice relaxation,$\Delta H^{i}_{r}$, which follows the transfer of the O ion to the vacancy site. In (b), the case of electron mobility is shown. Here, the shaded circles are Ce$^{4+}$ sites well away from the O vacancy site. The empty circle is a Ce$^{3+}$ site which is assumed to belong to the vacancy cluster. All Ce sites belong to the O$_{h}$ point group. The potential experienced by an electron as it hops to the Ce$^{4+}$ sites is shown by the dotted curve. The Ce$^{3+}$ site is lower in energy by $\Delta H^{e}_{ass}$. The potential that results from the lattice relaxation after the electron has transferred to the middle Ce site is shown by the solid curve. This illustration shows that there is a trapping potential acting on the Ce$^{3+}$ site due to its proximity to the O vacancy.}%
\label{phenomenology}%
\end{center}
\end{figure}

We define a vacancy cluster consisting of the vacancy together with all the atoms up to and including the nnn Ce sites. Ionic conductivity involves the migration of the vacancy cluster on the O sublattice. The occupied O-lattice sites (so called normal sites) and the vacancy sites are assumed to be energetically inequivalent with the difference in energy given by $\Delta H^{i}_{ass}$, Fig. \ref{phenomenology}(a). Similarly, to describe electronic hopping on the cationic sublattice, we assume two types of Ce sites - the Ce$^{3+}$ and Ce$^{4+}$ which are energetically inequivalent. The energy difference between these sites is $\Delta H^{e}_{ass}$, Fig. \ref{phenomenology}(b).

 To mobilize a charge carrier requires that its associated energy barrier be overcome, so an energy of at least $\Delta H^{e}_{ass}$, for electrons, or $\Delta H^{i}_{ass}$, for ions, must be provided by thermal fluctuations.  In the case of ionic conductivity there is an additional barrier energy, $\Delta H^{i}_{b}$, associated with O$^{2-}$ jumping to the saddle point at the boundary of the first coordination shell of Ce ions. Once the O$^{2-}$ reaches the vacancy site, the lattice relaxes to the new configuration. The lattice relaxation has an associated energy, $\Delta H^{i}_{r}$. Thus the mobility barrier of an O ion is given by the sum: $\Delta H^{i}_{m}=\Delta H^{i}_{b}+\Delta H^{i}_{r}$.  For electronic conductivity, no barrier energy analogous to $\Delta H^{i}_{b}$ exists so only the energy $\Delta H^{e}_{ass}$ need be provided for the electron can tunnel to a neigbouring Ce site. However, as in the ionic case, once the electron transfer has occurred, the lattice relaxes with an energy given by $\Delta H^{e}_{r}$. Consequently, the mobility energy of an electron is given by: $\Delta H^{e}_{m}=\Delta H^{e}_{r}$.

For a mixed conductor with several charge carriers, the total conductivity is given by:
\begin{equation}
\sigma =\sum_{j}\sigma_{j} = \sum_{j}n_{j}q_{j}\mu_{j} \label{gdenough_0}
\end{equation}
where $\sigma_{j}$, $n_{j}$, $q_{j}$ and $\mu_{j}$ refer to the conductivity, density, charge and mobility of the $j$th species of charge carrier. The O vacancies and/or electrons are bound to the O vacancy cluster so to obtain `free' charge carriers, energy corresponding to at least $\Delta H^{i}_{ass}$ and $\Delta H^{e}_{ass}$ must be provided to the system for vacancies and electrons respectively. The thermally activated free carrier population is given by
\begin{equation}
n_{j} = n^{j}_{0}\exp\left(\frac{-\Delta H^{j}_{ass}}{k_{B}T}\right)  \label{gdenough_1a}
\end{equation}
where $k_{B}$ is the Boltzmann constant and $T$ is the temperature. For itinerant electrons, the mobility is given by:
\begin{equation}
\mu_{e} =\frac{e\tau_{s}}{m^{*}}  \label{gdenough_1}
\end{equation}
where $e$ and $m^{*}$ are the electron charge and mass and $\tau_{s}$ is the mean free time for electron scattering. For SPM and ionic conductivity, the mobility is diffusive and activated. It is related to the diffusivity, $D_{j}$, by the Nernst-Einstein relation \cite{Mehrer2007}:
\begin{equation}
\mu_{j} =\frac{q_{j}D_{j}}{k_{B}T}  \label{gdenough_2}
\end{equation}
A general expression for the diffusion coefficient,$D_{j}$, is obtained from random walk theory which assumes \emph{uncorrelated} motion of the charge carriers. It is found that this expression has the same form for both SPM electronic \cite{Tuller1977} and ionic \cite{Goodenough1984,Goodenough2003,Inaba1996} conductivity given by:
\begin{equation}
D_{j} =\gamma_{j} (1-c_{j})a^{2}_{j}\nu^{j}_{0}\exp\left(\frac{-\Delta H^{j}_{m}}{k_{B}T}\right)  \label{gdenough_2a}
\end{equation}
where the quantity, $\gamma_{j}$, has a different meaning between electronic and ionic conductivity as will be described below. The quantity, $c_{j}$, is the concentration of normal sites, i.e., sites which are occupied by O atoms for ionic conductivity, and by Ce$^{3+}$ ions for SPM conductivity. Hence $c_{j} = \frac{\theta_{j}}{N_{j}}$ with $\theta_{j}$ and $N_{j}$ being the numbers of occupied and total available sites of type $j$ respectively. The attempt frequency, $\nu^{j}_{0}$ is reported to be $\sim 10^{12}s^{-1}$ for ionic conductivity \cite{Goodenough1984} and $\sim 10^{13}s^{-1}$ for SPM electronic conductivity \cite{Tuller1977} which corresponds to the optical mode vibration frequencies.

Finally, the quantity, $\gamma_{j}$, for the case of ionic conductivity is given by:
\begin{equation}
\gamma_{j} =\frac{1}{6}z\exp\left(\frac{\Delta S^{j}_{m}}{k}\right)  \label{gdenough_2b}
\end{equation}
with $z$ being the coordination number in the O sublattice of the site from which the O ion hops. For slightly reduced CeO$_{2}$, $z$ has the same value as it does in the unreduced phase so that $z=6$. For the electronic case, the value of $\gamma_{j}$ depends on whether the electron transfer is adiabatic or non-adiabatic\cite{Holstein1959,Emin1969,Tuller1977}:
\begin{equation}
   \gamma_{j} \sim \left\{
     \begin{array}{ll}
       1 & : \text{adiabatic electron transfer}\\
       \frac{t^{2}}{\hbar \sqrt{\lambda k_{B}T}} & : \text{non-adiabatic electron transfer}
     \end{array}
   \right.       \label{gdenough_2c}
\end{equation}
Here, $t$ is the electron hopping matrix matrix element and $\lambda$ is the reorganization energy.

For the motional free energy, we have already noted the energy terms that enter into the calculation of $\Delta H^{j}_{m}$ for both ionic and electronic conductivity. From Eqs. \ref{gdenough_2a}, \ref{gdenough_2}, \ref{gdenough_1a} and Eq. \ref{gdenough_0} we get:
\begin{equation}
\sigma =\sum_{j}\sigma_{j} = \sum_{j}\left(A_{j}/T\right)\exp\left(\frac{-E^{j}_{a}}{k_{B}T}\right) \label{gdenough_3}
\end{equation}
where,
\begin{equation}
E^{j}_{a} = \Delta H^{j}_{m} + \Delta H^{j}_{ass}  \label{gdenough_4}
\end{equation}
\begin{equation}
A_{j} = \gamma_{j}(q^{2}_{j}/k)c(1-c)a^{2}_{j}n^{j}_{0}\nu^{j}_{0}\exp\left(\frac{\Delta S^{j}_{m} + \Delta S^{j}_{ass}}{k}\right)  \label{gdenough_5}
\end{equation}
Small polaron (SP) electron transfer can be equivalently described in the Marcus-Hush theory which provides a general expression for the calculation of the rate of electron transfer \cite{Marcus1985}:
\begin{equation}
k_{et} = \frac{2\pi t^{2}}{\hbar \sqrt{4\pi k_{B}T \lambda}}exp\left(-\frac{\left(\Delta G^{0} + \lambda\right)^{2}}{4\lambda k_{B}T}\right)
\label{marcus01}
\end{equation}
where $k_{et}$ is the electron transfer rate or jump frequency ($s^{-1}$), and $\Delta G^{0}$ is the Gibbs free energy of the electron transfer reaction. This model of electron transfer enables us to calculate $t$ from some experimental data on mobility and activation energies.

The electron transfer rate, $k_{et}$, is related to the diffusion coefficient \cite{Mehrer2007a,Heitjans2005,Shewmon1989}:
\begin{equation}
D_{j} = \frac{zd^{2}k_{et}}{6}
\label{marcus01b}
\end{equation}
Here, $z$ is the number of equivalent nearest neighbour sites onto which the electron can hop and $d$ is the jump distance. As we only have experimental data of mobilities for slightly reduced ceria, we consider the Ce sublattice to be that of CeO$_{2}$ so that $z = 12$ and $d = a\sqrt{2}/2$, where $a$ is the lattice constant.

\begin{figure}
[bp]
\begin{center}
\includegraphics[width=0.99\textwidth]{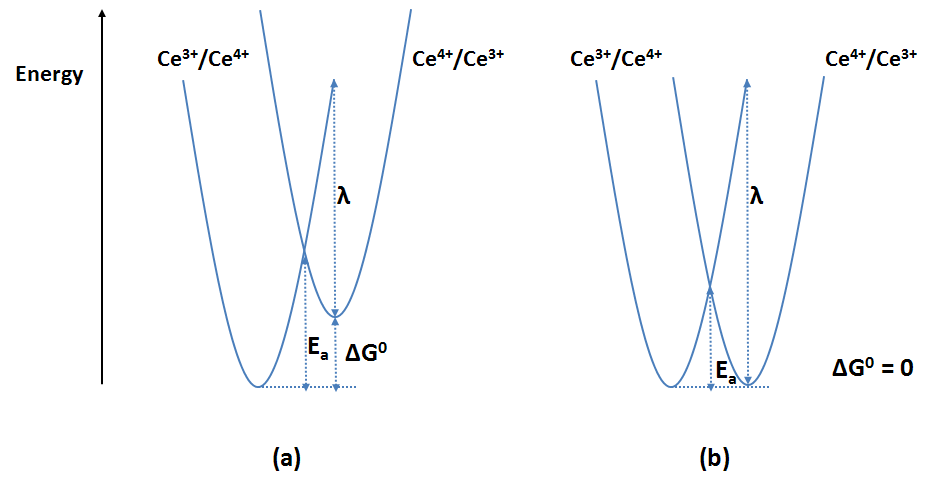}%
\caption{A schematic for the small polaron model in Marcus-Hush theory. (a) Electron is hopping away from a site which is part of the vacancy cluster. (b) Electron is hopping between sites well away from O vacancy sites. The energy parameters are $\lambda$ - the reorganization energy, $E_{a}$ - the activation energy and $\Delta G^{0}$ - the Gibbs free energy of the electron transfer process. The $\Delta G^{0} \sim 0$ for hopping between sites which are both well away from the O vacancy sites.}%
\label{marcus}%
\end{center}
\end{figure}

The activation energy and the reorganization energy are related to each other by:
\begin{equation}
 E_{a}= \frac{\left(\lambda+\Delta G^{0}\right)^{2}}{4 \lambda}
\label{marcus02}
\end{equation}
Fig. \ref{marcus} is a simple schematic of how SP transport in CeO$_{2-x}$ phases can be viewed in the Marcus-Hush picture (See also Fig. 7.8 in Mahan \cite{Mahan2000}).
For $\Delta G^{0} = 0$, from Eqs. \ref{marcus01}-\ref{marcus01b}, we have the relationship required to calculate the hopping matrix element from the mobility:
\begin{equation}
t = \sqrt{\frac{\hbar \left(k_{B}T\right)^{\frac{3}{2}}\left(\frac{\lambda}{\pi}\right)^{\frac{1}{2}}\mu}{ea^{2}}}\exp \left(\frac{\lambda}{8k_{B}T}\right)
\label{marcus01c}
\end{equation}
%
%
%
\subsection{Parameters for Conductivity in CeO$_{2-x}$ Phases}
We summarise in Table \ref{parameters} both experimental and theoretical results for the main parameters in the conductivity equation, Eq. \ref{gdenough_3}, in both the electronic and ionic cases.

\begin{table*}[htb] \centering
\caption{Some parameters for conductivity extracted from experimental and theory (Ref. \cite{Butler1983}) results in CeO$_{2-x}$ phases. $\Delta H^{e}_{m}$ and $\Delta H^{i}_{m}$ are the motional enthalpies for electronic and ionic conductivity respectively, $\Delta H^{e}_{ass}$ and $\Delta H^{i}_{ass}$ are the association energies for the electron and O vacancy respectively and $E_{a}$ is the sum of the respective energies. The energies are given in $\unit{\text{eV}}$.}%
\setcounter{mpfootnote}{\value{footnote}}
\begin{minipage}{\textwidth}
\renewcommand{\thempfootnote}{\arabic{mpfootnote}}
\begin{tabular*}{\textwidth}{@{\extracolsep{\fill}}{c} {c} {c} {c} {c} {c} {c} {c} }\hline\hline
\multicolumn {4}{c|}{Electronic} & \multicolumn {4}{c}{Ionic}\\\hline
$\Delta H^{e}_{m}$ & $\Delta H^{e}_{ass}$ & $E_{a}$ & Ref. & $\Delta H^{i}_{m}$ & $\Delta H^{i}_{ass}$ & $E_{a}$ & Ref.\\\hline
$0.40$ & $0.11$ & $0.51$ & \cite{Huang1998} & $0.63$ & $0.19$ & $0.82$ & \cite{Huang1998}\\
$-$ & $-$ & $0.40$ & \cite{Tuller1977} & $0.50$ & $-$ & $-$ & \cite{Fuda1984,Fuda1985,Butler1983}\\
$-$ & $-$ & $0.22$ & \cite{Blumenthal1974} & $0.61$ & $-$ & $-$ & \cite{Wang1981}\\
$-$ & $-$ & $0.22\footnotemark[1]$ & \cite{Lai2005} & $-$ & $-$ & $0.67\footnotemark[1]$ & \cite{Lai2005}\\
$-$ & $-$ & $0.25\footnotemark[2]$ & \cite{Steele2000} & $-$ & $-$ & $0.64\footnotemark[2]$ & \cite{Steele2000}\\
$-$ & $-$ & $0.52\footnotemark[3]$ & \cite{Wang2000} & $-$ & $-$ & $0.71\footnotemark[3]$ & \cite{Wang2000}\\
\\\hline\hline
\end{tabular*}
\footnotetext[1]{obtained for Sm-doped ceria: Ce$_{0.85}$Sm$_{0.15}$O$_{1.925-\delta}$ (SDC15)}
\footnotetext[2]{obtained for Gd-doped ceria: Ce$_{0.9}$Gd$_{0.1}$O$_{1.95-\delta}$ (GDC10)}
\footnotetext[3]{obtained for Gd-doped ceria: Ce$_{0.8}$Gd$_{0.2}$O$_{1.9-\delta}$ (GDC20)}
\setcounter{footnote}{\value{mpfootnote}}
\end{minipage}
\label{parameters}%
\end{table*}%
\subsection{Signatures of Hopping Conductivity}
Table \ref{parameters} shows that the motional enthalpy for electronic conductivity is $\geq 0.2\unit{\text{eV}}$. The fact that the electronic conductivity is activated is consistent with incoherent (hopping) transport associated with localized charge. However, the activation energy is much less than the energy gap from the $4f$ states to the delocalized $5d$ band in these materials which, e.g., in CeO$_{2}$ is $\sim 6\unit{\text{eV}}$ (compare Fig. \ref{bandstructures}).

SP transport can also be investigated by measurement of the Seebeck coefficient of a material. In this case, the temperature dependence of the Seebeck coefficient is used to determine whether or not hopping mobility exists. We have noticed that mainly two models have been used in the literature to describe the Seebeck coefficient of SP transport. The first one is more applicable to broad band semiconductors and shows a $1/T$ temperature dependence \cite{Salamon2004,Hundley1997}:
\begin{equation}
S = \left(\frac{k_{B}}{e}\right)\left(\alpha + \frac{E_{S}}{k_{B}T}\right)
\label{broadband}
\end{equation}
where $\alpha$ is a coefficient of order unity. Eq. \ref{broadband} is valid at low to moderate temperatures, i.e., $k_{B}T \leq E_{S}$. At high temperatures, the Heikes formula, which is temperature-independent should be used \cite{Heikes1961,Chaikin1976}:
\begin{equation}
S = \left(\frac{k_{B}}{e}\right)\ln\left(\frac{2(1-c)}{c}\right)
\label{narrowband}
\end{equation}
Chaikin and Beni argue that Eq. \ref{broadband} should not be used for narrow band materials as it does not describe localized states well \cite{Chaikin1976}.

\subsection{Decoupling the ionic from the electronic conductivity}
Both the Hall and Seebeck coefficients have the same sign as the charge carrier and so can be used to establish the type of charge carrier involved.
A negative Seebeck coefficient was reported by Tuller and Norwick \cite{Tuller1977} implying that the electrons are the majority carriers  for compositions where $0\leq x \leq0.24$ in CeO$_{2-x}$.

Another method to distinguish between electronic and ionic conductivities is impedance spectroscopy \cite{Lai2005}. Here, a Nyquist plot of the the impedance spectrum gives the so-called `half tear-drop' profile for a mixed conductor which arises from two charge carriers with different time constants being superimposed on the same curve. The results of Lai and Haile \cite{Lai2005} appear to corroborate the finding by Tuller and Norwick \cite{Tuller1977} that electrons are the majority carriers in slightly reduced ceria. However, comparison of the two sets of results is not a simple matter as the former researchers used Sm-doped ceria whereas the latter used slightly reduced ceria.

\section{Electronic Conductivity in Slightly Reduced CeO$_{2-x}$ Phases}
\label{Sec II}
Experimental work on electronic charge transport in the
CeO$_{2-x}$ phases has been interpreted in terms of the small polaron model \cite%
{Blumenthal1974,Tuller1977}. Tuller and Norwick found a reasonable fit to Eq. \ref{narrowband} for the Seebeck coefficient of the CeO$_{2-x}$ phases \cite{Tuller1977}. Within the small polaron description the propagation of the charge carrier through the lattice is associated
with a propagation of a lattice distortion.  There is no reason why the polaron must be confined to the vicinity of the oxygen vacancy.  It was shown in Ref. \cite{Shoko2009b} from the BVM analysis that most of the excess charge in the crystal of Ce$_{11}$O$_{20}$ is localized on next nearest neighbour Ce sites from the O vacancy. In the same report, a prediction was also made that all the excess charge in Ce$_{6}$O$_{11}$ should be localized on the next nearest neighbour Ce sites. Thus, we see that the charge distributions obtained from the
BVM do not contradict the SPM for small values of $x$ and we can make a connection between the two pictures.
The transport data can be used to obtain an estimate of the hopping matrix element, $t$, from Eq. \ref{marcus01c} using $a = 5.411\unit{\AA}$ for the lattice constant we find:
\begin{equation}
 t = 0.006\unit{\text{eV}}
\label{marcus05}
\end{equation}
An estimate of $t$ can also be obtained from the Harrison method (see Section \ref{Sec III}) which gives $t \sim 0.1 \unit{\text{eV}}$. The bandwidth of the $4f$ level in CeO$_{2}$, $W$, calculated from density functional theory (DFT) is $\sim 1.4\unit{\text{eV}}$ \cite{Shoko2009a,Hay2006}. The bandwidth is related to the hopping integral by $W = 16t$ so we have $t \sim 0.09\unit{\text{eV}}$ in agreement with the result from the Harrison method. In contrast, the value of $t$ in Eq. \ref{marcus05} appears too low. However, since $n_{e}(T)$ is also thermally activated, the mobility cannot be accurately determined from fits over a small temperature range.  Moreover, correlation effects might be important in reducing the effective mobility.

\section{Electronic Conductivity in CeO$_{2-x}$ Phases at Intermediate Reduction}
\label{Sec III}
We now consider the electronic conductivity in phases of intermediate reduction. We discuss the example of Ce$_{7}$%
O$_{12}$. Since the charge distribution obtained from the BVM suggests a delocalization of the charge on the Ce(2) sublattice \cite{Shoko2009,Shoko2009b}, we include calculations of the hopping matrix elements by the Harrison method \cite{Harrison2004a}. The results enable us to discuss the question of charge delocalization in this crystal in a somewhat quantitative way.

\begin{figure}[tbp]
\begin{center}
\includegraphics[width=0.6\textwidth]
{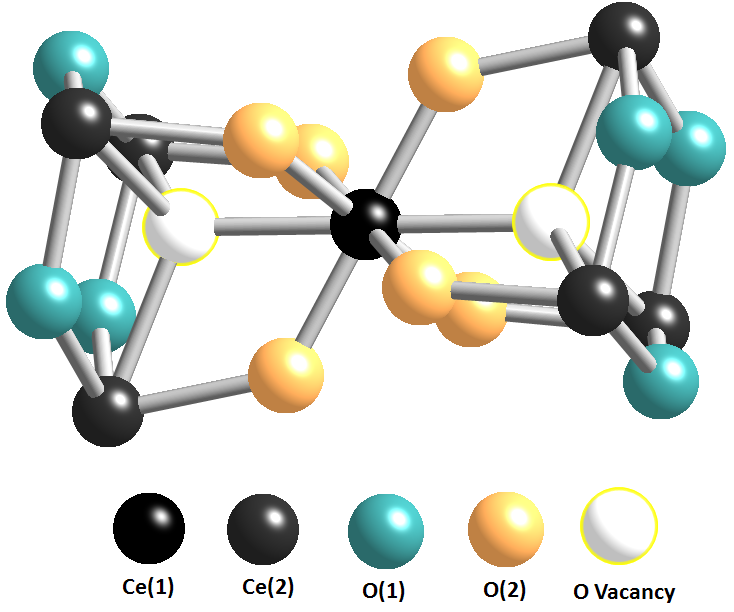}
\end{center}
\caption[Geometry of the Ce$_{7}$O$_{12}$ divacancy]{Geometry of the Ce$_{7}$%
O$_{12}$ divacancy showing the $S_{6}$ symmetry at the corner shared by the
two slightly distorted tetrahedra. The divacancy cluster shown contains all
the O atoms associated with the divacancy unit. Note that the Ce(2) sites
are coordinated by $7$ O atoms and therefore for each Ce(2) site, four Ce-O
bonds are not shown in this figure. The site symmetry of the Ce(1) site is $%
S_{6}$ while that of the Ce(2) sites is triclinic. Estimates of the site valences from the BVM give $+3.67$ and $+3.21$ for Ce(1) and Ce(2) respectively.}
\label{divacancy01}
\end{figure}

A key structural unit for discussing the electronic conductivity
in this crystal is the divacancy cluster shown in Fig. \ref{divacancy01}. It
results from the removal of two O atoms from the parent fluorite structure of CeO$_{2}$. Two distinct Ce sites
are obtained; the $S_{6}$ site with two nearest neighbour O vacancies and the $i$ site with only one O vacancy neighbour. The $S_{6}$ site has been
called the divacancy site \cite{Kang1997} and it forms a shared corner
between two coordination tetrahedra of the O vacancies as illustrated in
Fig. \ref{divacancy01}.
If we consider one of the O vacancies in the divacancy, the bond
valence results indicate that the charge delocalizes among the
Ce(2) sites that form an equilateral triangle. As different triangles are 
connected by relatively short Ce-O bonds, it is plausible that charge is
delocalized throughout the entire crystal on the Ce(2) sublattice.  If this is the
case, we can discuss the low temperature electronic conductivity of Ce$%
_{7}$O$_{12}$ based on a simplified picture of a system of connected
equilateral triangles each with a delocalized charge at one-third filling.
We can disregard the Ce(1) sites because the electrons from the O vacancy
do not occupy these sites. It is probably the case
that a large activation energy is associated with the hopping of an
electron onto Ce(1) sites. %
\begin{figure}
[ptb]
\begin{center}
\includegraphics[width=0.6\textwidth]{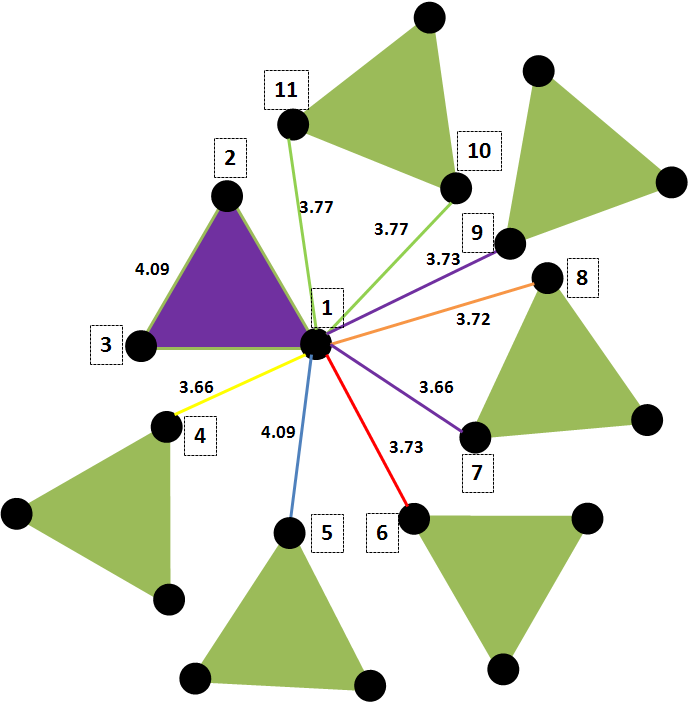}%
\caption{Schematic of part of the Ce(2)
sublattice of Ce$_{7}$O$_{12}$ showing the equilateral triangles connected
to the Ce$_{1}$ site. The boxed numbers are site labels which we reference
in the notation Ce$_{sitelabel}$. The numbers on the straight lines
connecting the Ce sites are the direct Ce-Ce distances. The equilateral
triangles are defined by joining together the three Ce(2) sites which are
nearest neighbours to the same O vacancy. In this schematic, the reference
triangle is highlighted in dark grey (purple) and the rest of the other
triangles (grey (green)) are nearest neighbours to the Ce$_{1}$ site on the
reference triangle. Each equilateral triangle contains two electrons
resulting from the formation of the O vacancy they coordinate. According to
the bond valence results of \textbf{I}, the two electrons are delocalized
among all three Ce(2) sites of the equilateral triangle. Thus, the three
Ce(2) sites coordinating the same O vacancy can be described as an
equilateral triangle at one-third filling.}%
\label{ce7o12_delocalize_3}%
\end{center}
\end{figure}
Fig. \ref{ce7o12_delocalize_3} shows a schematic of this simplified model where we
have also included the direct Ce-Ce distances to give an indication of the
relative separation between the equilateral triangles. We see that the
distances between the triangles range from $3.66\unit{}$ to $4.09\unit{%
\text{\AA}%
}$. The latter is the intra-triangle distance. Thus, inter-triangle
distances are shorter or equal to the intra-triangle distances and thus the inter-triangle electron hopping matrix
element may be comparable to the intra-triangle matrix element.

The direct Ce-Ce distances are summarised in Table \ref{Matrix} along with the matrix
elements for direct $f$-$f$ coupling between neighbouring Ce sites, $t_{ff}$. Table \ref{Matrix} shows that all but one of the matrix elements for
inter-triangle electron hopping are larger than the intra-triangle value of $%
0.01\unit{eV}$. The electron hopping matrix elements between lattice sites, $t_{ff}$
were calculated from Harrison's method of universal parameters \cite{Harrison2004a}.

\begin{figure}
[ptb]
\begin{center}
\includegraphics[width=0.6\textwidth]{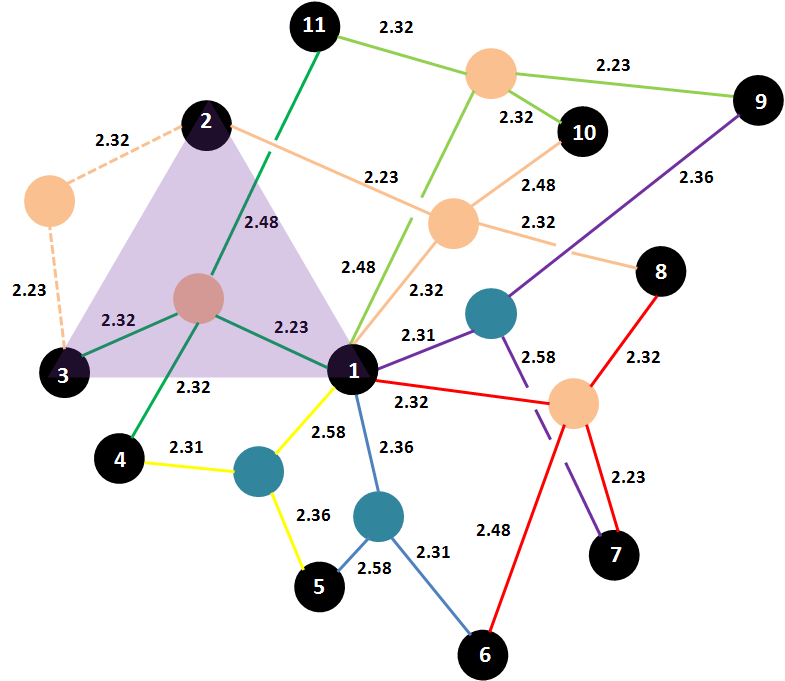}%
\caption{Schematic showing the Ce-O
bonds in the Ce(2) sublattice. Except in the case of the highlighted
triangle, only Ce-O bonds connecting the Ce$_{1}$ site to one of its ten
nearest neighbour Ce sites. The highlighted triangle consists of the three
Ce sites (including Ce$_{1}$) which are nearest neighbours to the same O
vacancy. There are two electrons delocalized in this triangle so that it can
be viewed as an equilateral triangle at one-third filling. The dotted Ce-O
bonds are not connected to the Ce1 site but have been included to show how
all the three Ce sites in the equilateral triangle are connected by Ce-O
bonds. The white numbers on the Ce sites are site labels which correspond to
those in Fig. \protect\ref{ce7o12_delocalize_4} whereas the numbers shown on
the lines connecting the various sites are the Ce-O bond lengths in $\unit{\text{\AA}}$. The bond lengths are summarised in Table \protect\ref{Matrix}. All the
Ce sites in the Ce(2) sublattice are equivalent and each is
seven-coordinated to O atoms and belongs to an equilateral triangle where
the charge is delocalized as shown for Ce$_{1}$. The identification of the
sites is as follows: black circles - Ce, grey (teal) - O(1) and light grey
(orange) - O(2).}%
\label{ce7o12_delocalize_4}%
\end{center}
\end{figure}

Since electrons may become delocalized on the Ce lattice by indirect hopping via oxygen sites we also should consider relevant Ce-O bond distances.  Fig. \ref{ce7o12_delocalize_4} shows the Ce-O bond lengths associated with
the bonding of the Ce$_{1}$ site to its ten nearest neighbour Ce sites. The Ce-O bonds are summarized in Table \ref{Matrix} together
with the matrix elements for electron hopping between Ce sites via an O
site, i.e., $f$-$p$-$f$ hopping, $t_{eff}$.  Indirect hopping between
two Ce(2) sites via an intervening O site involves a two-step process in which the electron first hops from the first
Ce(2) to the O site and then from the O site to the second Ce(2) site. If $%
t_{fp}$ is the matrix element for hopping between a Ce(2) site and an O
site, the overall matrix element for the electron hopping between two Ce(2)
sites via an O site, $t_{eff}$, is then given by Eq. (\ref{hopping}):
\begin{equation}
t_{eff}=\frac{t_{fp}^{2}}{\varepsilon _{f}-\varepsilon _{p}}  \label{hopping}
\end{equation}%
where $\varepsilon _{f}-\varepsilon _{p}$ is the energy gap between the Ce $%
4f$- and the O $2p$ levels, which, in the calculation, we assume to be $\sim
2\unit{eV}$. The matrix elements for indirect electron hopping given in
Table \ref{Matrix} are calculated from this formula \cite%
{Harrison2004a,PhysRevB.36.2695,McConnell1961}. We note that all the matrix elements
listed in Table \ref{Matrix} are for the $\sigma $-$\sigma $ interaction
between the two orbitals as they are the most favourable.
From Table \ref{Matrix}, we see that the $t_{eff}$ matrix elements for
intra-triangle hopping are equal and have the value $0.15\unit{eV}$.
The same value is obtained for electron
hopping from the Ce$_{1}$ site to the Ce$_{4}$ and Ce$_{7}$ sites. An electron at Ce$_{1}$ has an equal probability of hopping
onto sites Ce$_{2}$ and Ce$_{3}$ (intra-triangle hopping) or onto sites Ce%
$_{4}$ and Ce$_{7}$ (inter-triangle hopping). The matrix elements for an
electron at Ce$_{1}$ to hop to Ce$_{8}$ are also similar. If these
matrix elements are a fair indicator of the tendency towards charge
delocalization, then they support our conjecture
that the charge should be delocalized throughout the Ce(2) sublattice of Ce$%
_{7}$O$_{12}$. This delocalization of charge throughout the crystal would
imply that Ce$_{7}$O$_{12}$ should exhibit metallic conductivity at low
temperature. If this is true, then Ce$_{7}$O$_{12}$ should have the highest low-temperature electronic conductivity of all the reduced phases of ceria. 
  
We have not found any reports of low temperature electronic
conductivity measurements of Ce$_{7}$O$_{12}$.
\begin{table}[tbp] \centering%
\caption{A summary of the Ce-O bond lengths from Fig. \ref{ce7o12_delocalize_4} and the direct Ce-Ce distances from Fig. \ref{ce7o12_delocalize_3}. Matrix elements
for electron hopping calculated according to the Harrison method of universal parameters are included. The matrix
element, ${t_{eff}}$, refers to electron hopping between the Ce$_{1}$ site and its eleven nearest neighbour Ce sites via an O site whereas
${t_{ff}}$ refers to a direct hopping by $f$-$f$ coupling}%
\begin{tabular}{|l|l|l|p{2.4cm}|l|p{2.4cm}|l|}
\hline
Ce$_{i}$ & Ce$_{1}$-O, $\unit{%
\text{\AA}%
}$ & O-Ce$_{i}$, $\unit{%
\text{\AA}%
}$ & Ce$_{1}$-O-Ce$_{i}$ distance, $\unit{%
\text{\AA}%
}$ & $t_{eff}$, $\unit{eV}$ & Ce$_{1}$-Ce$_{i}$ distance, $\unit{%
\text{\AA}%
}$ & $t_{ff}$, $\unit{eV}$ \\ \hline
Ce$_{2}$ & $2.32$ & $2.23$ & $4.55$ & $0.15$ & $4.09$ & $0.01$ \\
Ce$_{3}$ & $2.23$ & $2.32$ & $4.55$ & $0.15$ & $4.09$ & $0.01$ \\
Ce$_{4}$ & $2.23$ & $2.32$ & $4.55$ & $0.15$ & $3.66$ & $0.03$ \\
Ce$_{4}$ & $2.58$ & $2.31$ & $4.89$ & $0.08$ & $-$ & $-$ \\
Ce$_{5}$ & $2.58$ & $2.36$ & $4.94$ & $0.07$ & $4.09$ & $0.01$ \\
Ce$_{5}$ & $2.36$ & $2.58$ & $4.94$ & $0.07$ & $-$ & $-$ \\
Ce$_{6}$ & $2.36$ & $2.31$ & $4.67$ & $0.12$ & $3.73$ & $0.03$ \\
Ce$_{6}$ & $2.32$ & $2.48$ & $4.80$ & $0.09$ & $-$ & $-$ \\
Ce$_{7}$ & $2.32$ & $2.23$ & $4.55$ & $0.15$ & $3.66$ & $0.03$ \\
Ce$_{7}$ & $2.31$ & $2.58$ & $4.89$ & $0.08$ & $-$ & $-$ \\
Ce$_{8}$ & $2.32$ & $2.32$ & $4.64$ & $0.13$ & $3.72$ & $0.03$ \\
Ce$_{8}$ & $2.32$ & $2.32$ & $4.64$ & $0.13$ & $-$ & $-$ \\
Ce$_{9}$ & $2.31$ & $2.36$ & $4.67$ & $0.12$ & $3.73$ & $0.03$ \\
Ce$_{9}$ & $2.48$ & $2.23$ & $4.71$ & $0.11$ & $-$ & $-$ \\
Ce$_{10}$ & $2.32$ & $2.48$ & $4.80$ & $0.09$ & $3.77$ & $0.02$ \\
Ce$_{10}$ & $2.48$ & $2.32$ & $4.80$ & $0.09$ & $-$ & $-$ \\
Ce$_{11}$ & $2.48$ & $2.32$ & $4.80$ & $0.09$ & $3.77$ & $0.02$ \\
Ce$_{11}$ & $2.23$ & $2.48$ & $4.71$ & $0.11$ & $-$ & $-$ \\ \hline
\end{tabular}%
\label{Matrix}%
\end{table}%
But some results for Pr$_{7}$O$_{12}$ and Tb$_{7}$O$_{12}$ may be applicable. In a study of the electrical
conductivities of PrO$_{2-x}$ and TbO$_{2-x}$, Rao \textit{et al.} \cite%
{Rao1970} found the highest conductivities for the compositions
corresponding to Pr$_{7}$O$_{12}$ and Tb$_{7}$O$_{12}$. The authors also
found that the Seebeck coefficient vanished at
this composition. It was suggested that the seven-coordinated cations with
an average charge of $+3.5$ would be responsible for the observed high
conductivity. We, however, emphasize here that the results of Rao \textit{et
al. }are for electronic conductivity at temperatures $\sim 200$-$600\unit{%
{}^{\circ}{\rm C}%
}$ whereas the bond valence results are for low
temperature. Electronic conductivity at high temperature may be dominated by
electron-phonon coupling effects which are absent at low temperature.

\begin{figure}
[ptb]
\begin{center}
\includegraphics[width=0.6\textwidth]{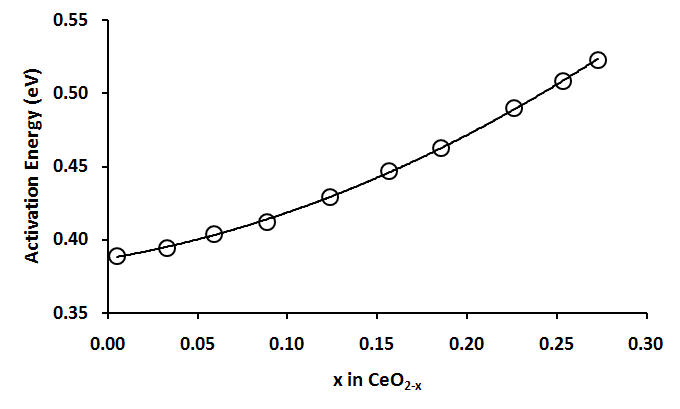}%
\caption{Variation of
activation energy from the high temperature region of the conductivity curve
($1000\unit{{}^{\circ}{\rm C}}$), with nonstoichiometry, $x$. Adapted from Tuller and Nowick \protect\cite%
{Tuller1977}.}%
\label{tuller}%
\end{center}
\end{figure}

We did not find any report on the electronic conductivity of reduced ceria
at the exact composition corresponding to Ce$_{7}$O$_{12}$. The results of
Tuller and Nowick \cite{Tuller1977} reproduced in Fig. \ref{tuller} are the
closest we got. %

The data in Fig. \ref{tuller} which support the SPM shows an
increase in the activation energy for the electron hopping with increasing $x
$ to about $x=0.25$ which is the upper limit of the composition range measured. An extrapolation of these results to the composition of Ce$_{7}
$O$_{12}$ ($x=0.29$) would suggest a higher activation energy for electron
hopping. However, it is not clear whether or not an
extrapolation is a valid extension in the interpretation of these data. As
the authors pointed out, it is expected that at this composition, an $n-p$
transition will occur which may invalidate any extrapolative interpretation.
Thus the situation regarding Ce$_{7}$O$_{12}$ appears somewhat inconclusive
from the data on the SPM we were able to find. That low temperature metallic
conductivity (i.e. band transport) should occur in Ce$_{7}$O$_{12}$ is unexpected from a
consideration of the electronic structures of CeO$_{2}$ and Ce$_{2}$O$_{3}$.
These two phases are relatively well characterized and form the boundaries
of the CeO$_{2-x}$ composition range. On this composition range, Ce$_{7}$O$%
_{12}$ is closer to Ce$_{2}$O$_{3}$ than it is to CeO$_{2}$. CeO$_{2}$ is a band (charge transfer) insulator whereas Ce$%
_{2}$O$_{3}$ is a Mott insulator. The crossover from incoherent (hopping) to coherent (band) transport will occur at a temperature such that $\delta (T^{*}) > e^{2}a/\hbar$.

\section{Results from Atomistic Models for Ionic Conductivity in CeO$_{2-x}$ Phases}
\label{Sec IV}
There is a large body of literature on the study of ionic
conductivity in doped ceria \cite%
{Trovarelli2002a,Inaba1996,Kharton2001,Yokokawa2006,Omar2006,Andersson2006,Omar2008,Yan2008}%
. It is known that the local crystal environment of the
O vacancies affects their mobility,  so defect cluster models that simulate the local environment  of the defect are employed.  Below, we review some key results of the literature and consider whether they are consistent with the BVM.

The mechanism of bulk ionic conductivity in doped or reduced ceria can be viewed either as oxygen self-diffusion or
vacancy diffusion (of course, oxygen atoms move in the opposite direction to oxygen vacancies) under an applied electric field. The ionic
conductivity is proportional to the product of the charge and concentration of the carriers and has an Arrhenius
temperature dependence with an activation energy \cite%
{Kilner1982,Kofstad1995,Mogensen2000}.

The process of doping CeO$_{2}$ with a trivalent oxide, M$_{2}$O$_{3}$, can
be described by:%
\begin{equation}
\text{M}_{2}\text{O}_{3}\rightleftharpoons 2\text{M}_{\text{Ce}}^{\prime }+%
\text{ }3\text{O}_{\text{O}}+V_{O}^{\cdot \cdot }  \label{doping}
\end{equation}%
where M$_{\text{Ce}}^{\prime }$ is a dopant ion on a Ce$^{4+}$ site. Note
that for M=Ce, Eq. \ref%
{doping} is equivalent to reduction of CeO$_{2}$, Eq. \ref{reduce}. For free O vacancies, the conductivity
at low temperature is described by Eq. \ref{gdenough_4} with $\Delta H^{j}_{ass} = \Delta S^{j}_{ass} = 0$, so the ionic conductivity is a
linear function of the O vacancy (and hence dopant) concentration provided that $\Delta S^{j}_{m}$
and $E^{j}_{a}$ are approximately constant. However, experimental observations show that both $E_{a}$ and $%
\sigma $ are highly nonlinear functions of the dopant concentrations, which means that a concentration dependence of $E^{j}_{a}$ should be assumed if Eq. \ref%
{gdenough_4} is to be used.

To explain the relationship between $E^{j}_{a}$ and dopant
concentration, O vacancy cluster models were proposed. In these so-called defect associate models, the O vacancies are
bound to the dopant ions and form stable associates of the the form $\left(
\text{M}_{Ce}^{\prime }V_{O}^{\cdot \cdot }\right) ^{\cdot }$ and $\left(
\left( \text{M}_{Ce}^{\prime }\right) _{2}V_{O}^{\cdot \cdot }\right) ^{x}$.
The latter is a neutral cluster which is only relevant at high dopant
concentrations. Note that the superscript $x$ is the standard label for this type of latter vacancy cluster
, it should not be confused with the $x$ in Eq. \ref{reduce} (which never appears in superscript). In order for an O vacancy to contribute to the ionic conductivity, it must be ionized to give a free O vacancy:%
\begin{equation}
\left( \text{M}_{Ce}^{\prime }V_{O}^{\cdot \cdot }\right) ^{\cdot
}\rightleftharpoons \text{M}_{Ce}^{\prime }+V_{O}^{\cdot \cdot }\text{ \ \
or }\left( \left( \text{M}_{Ce}^{\prime }\right) _{2}V_{O}^{\cdot \cdot
}\right) ^{x}\rightleftharpoons 2\text{M}_{Ce}^{\prime }+V_{O}^{\cdot \cdot }
\label{ionize}
\end{equation}%
The ionization of an O vacancy defect cluster costs an energy $\Delta H^{j}_{ass}$, and entropy $\Delta S^{j}_{ass}$ so that the
ionic conductivity in these clusters is described by Eq. \ref{gdenough_4} with non-zero enthalpy and entropy of association.
$\Delta H^{j}_{ass}$ is the association or binding energy between the O vacancy
and the dopant ion, M$_{Ce}^{\prime }$. Some values are given in Table \ref{parameters}. Ionic conductivity measurements give
information about $\Delta H^{j}_{ass}$ but no information about the types of
defect clusters in the sample or their local atomic arrangement. To
obtain this detailed information, various techniques have been adopted. Most results describing the
binding energies of the O vacancies to the dopant sites have come from
atomistic modelling \cite{Kilner1982,Inaba1996,Minervini1999,Ye2008,Wei2009}%
. Our discussion here is limited to low temperatures (up to $\sim 600\unit{K%
}$) where the formation of defect clusters is almost complete \cite%
{Inaba1996,Faber1989}.

\begin{figure}
[ptb]
\begin{center}
\includegraphics[width=0.6\textwidth]{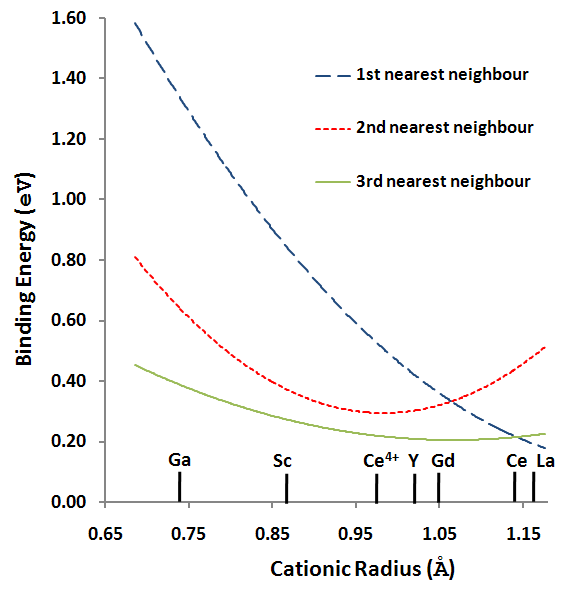}%
\caption{Trends in the binding energies
(from atomistic simulations) of an O vacancy to a dopant ion located in one
of the first, second or third coordination shell plotted as a function of
the \ dopant ionic radius. The binding energy is calculated by subtracting
the energy of the defect cluster from the sum of the energies of the
individual components. Thus a positive binding energy indicates a favourable
configuration for the defect cluster. Adapted from Minervini \textit{et al.}
\protect\cite{Minervini1999}.}%
\label{minervini_a}%
\end{center}
\end{figure}

\begin{figure}
[ptb]
\begin{center}
\includegraphics[width=0.99\textwidth]{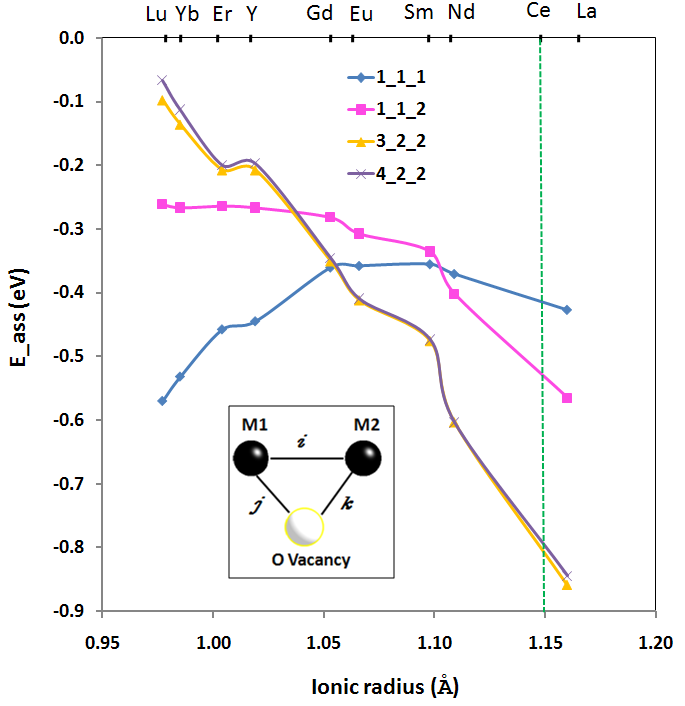}%
\caption{The binding energies of various
vacancy cluster configurations involving two dopant ions. The notation used
for the defect cluster geometry is $i\_j\_k$ where if the two dopants are
designated as M1 and M2, then M1 and M2 are $i$th nearest neighbours (nn)
and M1 and M2 are the $j$th and $k$th nn to the O vacancy respectively (See insert in Figure.). The
binding energy in this work is obtained by subtracting the total energy of
the supercell with the maximum energy from that of a particular supercell.
By this definition, a negative binding energy represents a favourable
structure. We have indicated the position of Ce$^{3+}$ in the figure based
on its ionic radius. The results suggest that based on this interpolation
for the Ce$^{3+}$ ion, \ configurations 3\_2\_2 and 4\_2\_2 are the most
favourable for arranging two Ce$^{3+}$ ions around an O vacancy. This is consistent with our BVM analysis of Ce$_{11}$O$_{20}$. Adapted
from Wei et al. \protect\cite{Wei2009}.}%
\label{wei_09}%
\end{center}
\end{figure}

$\Delta H^{j}_{ass}$ can be calculated for a particular geometry of the defect
associates and the results compared to the experimental $E_{a}^{\prime }$.
An important early experimental finding was that the ionic conductivity of doped ceria depended strongly
on the dopant ionic radius \cite{Gerhardt-Anderson1981}. The local atomic arrangement in a defect cluster
(and hence $\Delta H^{j}_{ass}$) also depends on the dopant ionic radius \cite%
{Butler1983,Faber1989}. Fig. \ref{minervini_a} shows the
trends in the binding energies calculated from atomistic modelling\cite%
{Minervini1999}. The most favourable arrangement in
the $\left( \text{Ce}_{Ce}^{\prime }V_{O}^{\cdot \cdot }\right) ^{\cdot }$
cluster has the Ce$^{3+}$ ions in the second coordination shell which is
what we found for Ce$_{11}$O$_{20}$ and what we expect for Ce$_{6}$O$_{11}$.
A similar study was recently done by Wei et al.\cite{Wei2009} and
the relevant part of their results is summarised in Fig. \ref{wei_09}. %

The results of Fig. \ref{wei_09} are for the composition Ce$_{30}$%
M$_{2}$O$_{63}$ which for M=Ce$^{3+}$ is the supercell most often used in DFT studies \cite%
{Skorodumova2001,Skorodumova2002,Fabris2005,Castleton2007,Andersson2007}.
 We have indicated in the figure where results of Ce$_{2}$O$_{3}$ are
expected to lie based on the ionic radius of Ce$^{3+}$. The results of Wei \textit{%
et al.} \cite{Wei2009} can thus be understood as corroborating those of
Minervini \textit{et al.} \cite{Minervini1999}. From atomistic simulations,
Pryde \textit{et al.} found that the binding energy for a vacancy cluster
with two Ce$^{3+}$ ions in the first coordination shell was
significantly less favourable than those with polarons in the second
coordination shell ($0.2\unit{eV}$ compared to $\sim 0.4\unit{eV%
}$)\cite{Pryde1995}. They also found that the defect
energetics of clusters involving In$^{3+}$ ions were qualitatively similar
to those for the Ce$^{3+}$ ions. The results of Pryde \textit{et al.} on
small polaron geometries are consistent with the BVM but contradict the standard picture. Similar results were
obtained for doped ZrO$_{2}$ by DFT methods \cite{Bogicevic2003}. We,
however, note that in contrast to these results, Deguchi et al. \cite%
{Deguchi2005} concluded from their extended x-ray absorption fine structure
(EXAFS) results that the dopant ions preferred to be in the first
coordination shell with respect to the O vacancy.

The higher binding
energies for defect associates in which the dopant ions are in the second
coordination shell have been explained in the following way \cite%
{Pryde1995,Minervini1999}: (i) the Coulomb interaction between the $%
V_{O}^{\cdot \cdot }$ and M$_{Ce}^{\prime }$ charged defects which favours
the first coordination shell for the dopant ions, (ii) lattice relaxation
which primarily has to do with the relaxation of the Ce$^{4+}$ with respect
to the $V_{O}^{\cdot \cdot }$ and M$_{Ce}^{\prime }$ charged defects around
it. Because of its large positive charge, the Ce$^{4+}$ ion prefers to relax
away from the $V_{O}^{\cdot \cdot }$ site towards the M$_{Ce}^{\prime }$
site. This mode of relaxation is not possible if the M$_{Ce}^{\prime }$ ion
is a nearest neighbour to the $V_{O}^{\cdot \cdot }$ site. Thus, this mode
of lattice relaxation always favours the second nearest neighbour site for
the dopant ion and (iii) a component of lattice relaxation driven by
ion-size effects which favour the nearest neighbour position for small
dopants. This is because the small ions prefer lower coordination. For M=Ce$%
^{3+}$, effect (ii) is the most dominant and an energy gain of $\sim 5\unit{%
eV}$ has been reported \cite{Minervini1999} which explains why the second
coordination shell is preferred for these ions. Although we have not
discussed the situation for high dopant concentrations where the neutral
trimer $\left( \left( \text{M}_{Ce}^{\prime }\right) _{2}V_{O}^{\cdot \cdot
}\right) ^{x}$ is expected to dominate, Minervini et al have shown that the
behaviour of these more complex systems is qualitatively similar to that of
the $\left( \text{M}_{Ce}^{\prime }V_{O}^{\cdot \cdot }\right) ^{\cdot }$
defect cluster \cite{Minervini1999}.

We note that the atomistic models use the generalized Mott-Littleton method
\cite{Mott1938} which assumes an ionic description of the crystal lattice and
the shell model for the ionic polarizabilities of the ions \cite%
{B.G.Dick1958}. It is possible that the assumption of an ionic
crystal for CeO$_{2}$ might not be consistent with the results of bond
valence calculations. Also, the models do not include
vacancy-vacancy interactions which might be important at high vacancy concentrations as reported recently
by Pietrucci \textit{et al}. \cite{Pietrucci2008}.

Additional data are needed for a comprehensive comparison with bond valence results. Ideally, one would also like curves of the binding energy
as a function of the dopant concentration.  The dopant of interest is Ce$_{2}$O$_{3}$ and since the
reduced phases of ceria considered in this paper are Ce$_{7}$O$_{12}$ ($\rightarrow 3$%
CeO$_{2}+2 $Ce$_{2}$O$_{3}$), Ce$_{11}$O$_{20}$ ($\rightarrow 7$CeO$_{2}+2$Ce$_{2}$O$%
_{3}$) and Ce$_{6}$O$_{11}$ ($\rightarrow 4$CeO$_{2}+$Ce$_{2}$O$_{3}$) the mole
fraction of the dopant will need to cover the range $0.4$ to $0.2$. These
are well within the capabilities of atomistic modelling. The range could be
extended to lower concentrations using the $2\times
2\times 2$ supercell typical of DFT calculations
and also in Fig. \ref{wei_09}%
. This classic case is interesting as it will provide an opportunity to
compare the results from ab initio electronic structure calculations with
those from atomistic modelling.

\section{Conclusion}

\label{Sec VI} For slightly reduced ceria, the charge distribution obtained
from the BVM is consistent with the description of the electronic
conductivity by the SPM. For the CeO$_{2-x}$ phases in the neighbourhood of $%
x=0.3$, we anticipate low temperature metallic conductivity or the
highest low temperature electronic conductivity of all the reduced ceria phases. This conclusion relies
on the estimation of electron hopping amplitudes from Harrison's method of
universal parameters. The detection of this metallic conductivity will discriminate between the standard picture and the
BVM desriptions of charge distribution in CeO$_{2-x}$ phases. We have also
considered the compatibility of the BVM charge distribution and the
atomistic models used to describe ionic conductivity in these phases. We
found that the atomistic models and the BVM give a consistent picture of the
location of the charge. However, the data are quite limited and it is
thus not possible to make a comprehensive assessment.

\section{Acknowledgements}

We are grateful to Prof. C. Stampfl at the University of Sydney for
introducing us to the field of cerium oxides. One of us (E. S) is
grateful to the Australian Commonwealth Government Department of Science
Education and Training for the award of the International Postgraduate
Research Scholarship (IPRS) and to the University of Queensland for the
University of Queensland International Postgraduate Research Scholarship
(UQIPRS). E. S is also grateful for the School of Mathematics and Physics
Postgraduate Travel Scholarship and the Australian Research Council
Nanotechnology Network (ARCNN). This work was also supported by the
Australian Research Council.
\bibliographystyle{elsarticle-num}
\bibliography{shoko_cpc_2009}

\section{Graphic Abstract}

Is low temperature metallic conductivity possible in Ce$_{7}$O$_{12}$? We argue that since the distribution of the excess charge in this crystal can be described by a simple model of a network of equilateral triangles each at $\frac{1}{3}$ filling, band conductivity may be expected.
\begin{figure}
[ptb]
\begin{center}
\includegraphics[width=0.6\textwidth]{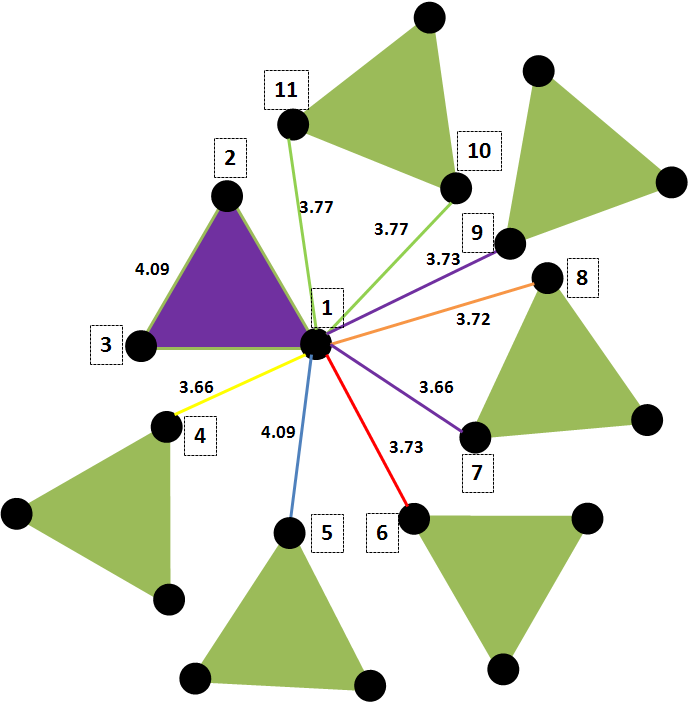}%
\label{ce7o12_delocalize_3a}%
\end{center}
\end{figure}
\end{document}